%% file: ICASSP_camera_ready.tex
\documentclass{article}
\usepackage{spconf,amsmath,amsthm,amsfonts,amssymb,mathtools,graphicx,subfigure,hyperref}
\usepackage[table]{xcolor}
\usepackage{colortbl}
\usepackage{url}            
\usepackage{booktabs}       
\usepackage{multirow}       
\usepackage{nicefrac}       
\usepackage{microtype}      
\usepackage{xcolor}         
\usepackage{rotating}
\usepackage{pifont}
\usepackage{tikz}
\usepackage{tkz-euclide}
\usepackage{bm}
\usetikzlibrary{shapes,arrows,arrows.meta}
\input{macros/tikz_init}
\input{macros/mathmacros}
\input{macros/tu_bs_colors}

\def\x{{\mathbf x}}

\renewcommand{\l}[1]{\small\tt{#1}}

\newcommand{\hskipt}{\hskip 0.32cm}

\definecolor{TUred}{RGB}{190,30,60} 
\definecolor{TUblue}{RGB}{102,180,211} 
\definecolor{mycolor1}{rgb}{0,0,0.8} 
\definecolor{mycolor2}{rgb}{0,0,0.8} 
\definecolor{mycolor3}{rgb}{0.8,0,0} 
\definecolor{mycolor4}{rgb}{0.8,0,0} 
\definecolor{mycolor5}{rgb}{0,0.6,0.0} 
\definecolor{mycolor6}{rgb}{0,0.6,0.0} 
\definecolor{mycolor6}{rgb}{0,0.35,0.0} 
\definecolor{mycolor7}{rgb}{0.9,0.5,0.0} 
\definecolor{mycolor8}{rgb}{0.9,0.5,0.0} 
\definecolor{mycolor9}{rgb}{0.9,0.9,0}
\definecolor{mycolor10}{rgb}{0.75,0.75,0}
\definecolor{mycolor11}{rgb}{0.4,0.4,0.4}
\definecolor{mycolor12}{rgb}{0.25,0.25,0.25}
\definecolor{mycolor13}{rgb}{0.4,0.4,0.4}
\definecolor{mycolor14}{rgb}{0.25,0.25,0.25}
\def\x{{\mathbf x}}

\def\x{{\mathbf x}}

\title{Noise-Robust AV-ASR Using Visual Features \\ Both in the Whisper Encoder and Decoder}
\name{Zhengyang Li, Thomas Graave, Björn Möller, Zehang Wu, Matthias Franz, Tim Fingscheidt}
\address{Technische Universit\"at Braunschweig\\
	Institute for Communications Technology\\
	Schleinitzstr. 22, 38106 Braunschweig, Germany}

\begin{document}
\ninept
\maketitle
\begin{abstract}
In audiovisual automatic speech recognition (AV-ASR) systems, information fusion of visual features in a pre-trained ASR has been proven as a promising method to improve noise robustness. In this work, based on the prominent \texttt{Whisper} ASR, first, we propose a simple and effective visual fusion method---\textit{use of visual features both in encoder and decoder (dual-use)}---to learn the audiovisual interactions in the encoder and to weigh modalities in the decoder. Second, we compare visual fusion methods in \texttt{Whisper} models of various sizes. Our proposed dual-use method shows consistent noise robustness improvement, e.g., a 35\% relative improvement (WER: 4.41\% vs.\ 6.83\%) based on \texttt{Whisper small}, and a 57\% relative improvement (WER: 4.07\% vs.\ 9.53\%) based on \texttt{Whisper medium}, compared to typical reference middle fusion in babble noise with a signal-to-noise ratio (SNR) of 0dB. Third, we conduct ablation studies examining the impact of various module designs and fusion options. Fine-tuned on 1929 hours of audiovisual data, our dual-use method using \texttt{Whisper medium} achieves 4.08\% (MUSAN babble noise) and 4.43\% (NoiseX babble noise) average WER across various SNRs, \textit{thereby establishing a new state-of-the-art in noisy conditions on the LRS3 AV-ASR benchmark.} Our code is at GitHub\footnote{\bf\url{https://github.com/ifnspaml/Dual-Use-AVASR}}. 
\end{abstract}
\begin{keywords}
audiovisual automatic speech recognition, information fusion, Whisper ASR, noise robustness
\end{keywords}
\section{Introduction}
\label{sec:intro}

The psychological finding that speech perception is inherently multimodal~\cite{Rosenblum2008} motivates researchers to build audiovisual automatic speech recognition (AV-ASR) systems, which leverage the speaker's lip movements to compensate for noisy speech. The noise robustness of AV-ASR systems promotes deployment in real-world such as automobiles~\cite{Wang2024} and smart glasses~\cite{Zmolikova2024}.

AV-ASR systems are usually comprised of one or multiple encoders for audiovisual feature extraction and an autoregressive decoder for next-token prediction. Recent research in AV-ASR systems~\cite{Rouditchenko2024,Rouditchenko2025,Kim2024,Yeo2025,Cappellazzo2025} developed rapidly by information fusion with pre-trained encoders or decoders such as \texttt{AV-HuBERT}, pre-trained by self-supervised learning (SSL) on unlabeled audiovisual data~\cite{Shi2022a,Shi2022,Li2025}, the encoder-decoder-based \texttt{Whisper} ASR trained by weakly supervised learning on 680k hours of pseudo-labeled speech data~\cite{Radford2023}, and decoders based on large language models (LLMs) like \texttt{Llama 3} pre-trained on over 15 trillion text tokens~\cite{Grattafiori2024} . 

The fusion methods in AV-ASR can be categorized as either early fusion~\cite{Shi2022,Shi2022a,Wang2024,Simic2024,Simic2025} in the encoder or middle fusion in the decoder~\cite{Receveur2016, Lohrenz2021a, Cappellazzo2024,Cappellazzo2025}. Early fusion is based on the fact that audio and video data are synchronized with a fixed frame rate ratio. Therefore, after simple temporal down/up-sampling, combination is realized by concatenation ~\cite{Shi2022,Shi2022a}, addition~\cite{Wang2024}, or with additional fusion modules~\cite{Simic2024,Simic2025,Kim2024}. In early fusion methods, the encoder already processes combined audiovisual features and effectively learns audiovisual interaction via its attention layers. However, early fusion becomes harder to train when applied to larger models, because the increased depth prevents effective gradient flow to the early fusion layers during backpropagation.

Middle fusion leverages separate encoders for audio (e.g., \texttt{Whisper} encoder~\cite{Radford2023}) and video (e.g., \texttt{AV-HuBERT} encoder), whose features are then processed in the decoder via fusion modules such as {\tt Flamingo} blocks~\cite{Alayrac2022,Rouditchenko2024}. The pre-trained decoder serves as a language model and decodes autoregressively conditioned on both audio and video features. Middle fusion achieves state-of-the-art performance in clean conditions~\cite{Cappellazzo2025,Rouditchenko2025} by exploiting the potential of advanced pre-trained decoders such as \texttt{Whisper} and an LLM decoder~\cite{Radford2023,Grattafiori2024}. However, unlike in early fusion, the audiovisual interactions are not modeled in middle fusion, leading to sub-optimal noise robustness.

In this work, first, we propose a dual use of visual features (dual-use) method to improve noise robustness of AV-ASR systems by applying visual features from \texttt{AV-HuBERT} in both \texttt{Whisper} encoder and decoder. Specifically, our first use is to inject visual features in the \texttt{Whisper} encoder with a zero-initialized smooth start, modeling audiovisual interactions during training. Our second use is inspired by previous work~\cite{Rouditchenko2024,Rouditchenko2025} feeding visual features to {\tt Flamingo} blocks~\cite{Alayrac2022}, which are integrated into the \texttt{Whisper} decoder, helping the decoder to balance modalities during decoding. Second, we compare our dual-use method with other fusion methods across various sizes of \texttt{Whisper} ASR. Compared to the early fusion method, our dual-use method improves noise robustness consistently in various sizes of \texttt{Whisper} ASR. In addition, our dual-use method significantly outperforms the middle fusion method~\cite{Rouditchenko2024,Rouditchenko2025} in noisy conditions with only a marginal increase in parameter count. Third, we investigate the influence of visual features from various \texttt{AV-HuBERT} blocks and various fine-tuning dataset sizes on noise robustness of AV-ASR, demonstrating that better visual features from later blocks improve noise robustness remarkably, and a larger fine-tuning dataset size improves the overall performance. Based on \texttt{Whisper medium}, our dual-use method achieves state-of-the-art performance on the LRS3 AV-ASR task.

The paper is structured as follows: In Section~\ref{sec:methods}, we introduce the baselines, reference fusion methods, and our proposed dual-use fusion method. In Section~\ref{sec:results}, following the experimental setup, we present results and discussion on the LRS3 AV-ASR task. The paper is concluded in Section~\ref{sec:conclusions}.

\section{Methods} \label{sec:methods}
\subsection{Baseline \texttt{Whisper} ASR Based on Acoustics}
\begin{figure}
	\centering
	\input{figure/whisper_based_avasr_with_avhubert_encoder.tikz}
	\caption{The audiovisual speech recognition system with \textbf{our proposed dual visual features use} in both the \texttt{Whisper} encoder and decoder.}\label{fig:dual-use}
\end{figure}
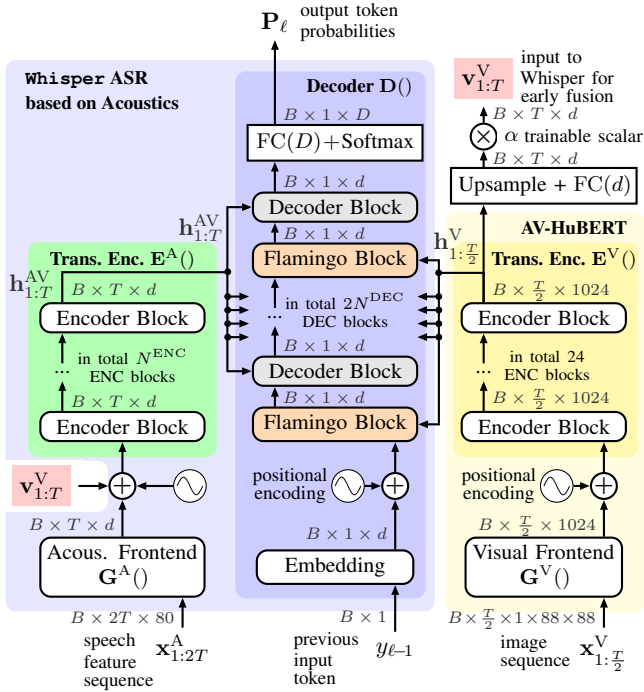

By training on 680k hours of speech data with weakly supervised learning, the transformer encoder-decoder-based \texttt{Whisper} ASR delivers advanced performance in clean and noisy conditions with audio-only input~\cite{Radford2023}. In this work, we choose \texttt{Whisper} ASR without and with fine-tuning on the target training data as baselines. In Fig.~\ref{fig:dual-use}, the encoder-decoder-based \texttt{Whisper} ASR is shown in light blue background (except the added visual features ${\VEC{v}}^\mathrm{V}_{1:T}$ in red background and inserted {\tt Flamingo} blocks in orange color). The {\tt Whisper} encoder process the input speech feature sequence $\VEC{x}_{1:2T}^\mathrm{A}=(\VEC{x}_{1}^\mathrm{A},\VEC{x}_{2}^\mathrm{A}, ...,\VEC{x}_{2T}^\mathrm{A})$ with serially connected acoustic frontend $\MAT{G}^\mathrm{A}()$ and transformer encoder $\VEC{E}^\mathrm{A}()$ in dark green, the latter comprising $N^\mathrm{ENC}$ transformer encoder blocks. The \texttt{Whisper} decoder with $N^\mathrm{DEC}$ transformer decoder blocks predicts the token probabilities $\mathbf{P}_\ell \in \mathbb{I}^D$ autoregressively, with $\mathbb{I}=[0,1]$, output sequence index $\ell \in \{1,..,L\}$, and a vocabulary size of $D$.

\subsection{Proposed Dual Visual Features Use for AV-ASR}
In Fig.~\ref{fig:dual-use}, we present our dual visual features use (dual-use) method. From the grayscale image sequence $\MAT{x}^\mathrm{V}_{1:\frac{T}{2}}=(\MAT{x}^\mathrm{V}_{1},\MAT{x}^\mathrm{V}_{2},...,\MAT{x}^\mathrm{V}_{\frac{T}{2}})$, the {\tt AV-HuBERT large} encoder (light yellow background), comprising a visual frontend $\MAT{G}^\mathrm{V}()$ and a transformer decoder $\MAT{E}^\mathrm{V}()$ with 24 serially stacked encoder blocks (dark yellow background), extracts visual latent representations $\VEC{h}^\mathrm{V}_{1:\frac{T}{2}}=(\VEC{h}^\mathrm{V}_{1},\VEC{h}^\mathrm{V}_{2},...,\VEC{h}^\mathrm{V}_{\frac{T}{2}})$. For their first use, we upsample the visual features $\VEC{h}^\mathrm{V}_{1:\frac{T}{2}}$ by frame-wise repetition, and then project them with a linear fully connected layer $\mathrm{FC}(d)$ to match the acoustic feature length $T$ and feature dimension $d$ of the \texttt{Whisper} model. The projected features are multiplied with a zero-initialized trainable scalar $\alpha$, resulting in visual features ${\VEC{v}}^\mathrm{V}_{1:T}=({\VEC{v}}^\mathrm{V}_{1},{\VEC{v}}^\mathrm{V}_{2},...,{\VEC{v}}^\mathrm{V}_{T})$ in red background. In the \texttt{Whisper} encoder, the acoustic features from the acoustic frontend $\MAT{G}^\mathrm{A}(\VEC{x}_{1:2T}^\mathrm{A})$ are added to the processed visual feature ${\VEC{v}}^\mathrm{V}_{1:T}$. Conditioned on the sum of both, the \texttt{Whisper} transformer encoder $\MAT{E}^\mathrm{A}()$ extracts audiovisual latent representations $\VEC{h}^\mathrm{AV}_{1:T}$ with modeling of audiovisual interactions:
\begin{equation}\label{eq:early_fusion}
	\VEC{h}^\mathrm{AV}_{1:T} = \VEC{E}^\mathrm{A}(\VEC{G}^{\mathrm{A}}(\VEC{x}^{\mathrm{A}}_{1:2T}) + \colorbox{red!20}{${\VEC{v}}^\mathrm{V}_{1:T}$})
\end{equation}
The second use of visual features $\VEC{h}^\mathrm{V}_{1:\frac{T}{2}}$ is by {\tt Flamingo} blocks (orange color)~\cite{Alayrac2022} integrated into the \texttt{Whisper} decoder (dark blue background). We follow the design detailed in~\cite{Rouditchenko2024,Rouditchenko2025}: The original \texttt{Whisper} decoder comprises $N^\mathrm{DEC}$ decoder blocks. Before each decoder block, one {\tt Flamingo} block is integrated, totaling to $2N^\mathrm{DEC}$ decoder blocks in the proposed decoder $\VEC{D}()$. Each {\tt Flamingo} block is built by serially connected multi-head cross attention and a feed-forward network, both with a gated mechanism with zero initialization~\cite{Alayrac2022}. The decoder $\VEC{D}()$ predicts token probabilities $\mathbf{P}_\ell$ autoregressively conditioned on the audiovisual latent representations $\VEC{h}^\mathrm{AV}_{1:T}$~(\ref{eq:early_fusion}) from our modified \texttt{Whisper} encoder, visual latent representations $\VEC{h}^\mathrm{V}_{1:\frac{T}{2}}$ from the \texttt{AV-HuBERT} encoder, and previous input tokens $y_{{1:\ell-1}}$, where earlier input tokens $y_{{1:\ell-2}}$ are implicitly saved as key-value cache in the transformer decoder, therefore not shown in Fig.~\ref{fig:dual-use}. The process can be written as:
\begin{equation}\label{eq:middle_fusion}
\mathbf{P}_\ell = \VEC{D}(\VEC{h}^\mathrm{AV}_{1:T}, \VEC{h}^\mathrm{V}_{1:\frac{T}{2}}, y_{{1:\ell-1}})
\end{equation}
With a search algorithm such as beam search on the token probabilities $\mathbf{P}_\ell$, a word sequence is decoded. The decoder $\VEC{D}()$, which is interpreted as a language model, can weigh features from these two encoders according to the context and noise conditions.

In this work, the proposed early fusion~(\ref{eq:early_fusion}) and the {\tt Flamingo}-based middle fusion~\cite{Rouditchenko2024,Rouditchenko2025} also serve as two stand-alone reference methods for comparison.

\section{Evaluation and Discussion}
\label{sec:results}

\subsection{Experimental Setup}
\textbf{Databases and pre-processing}: As training data, we use 433\,h Lip Reading Sentences 3 (LRS3) collected from TED and TEDx talks~\cite{Afouras2018b}, 224\,h Lip Reading Sentences 2 (LRS2) collected from British broadcasting corporation (BBC) programs~\cite{Chung2016}, and the 1326\,h pseudo-labeled English subset of the Voxceleb2 dataset~\cite{Chung2018}. Models are evaluated on the LRS3 AV-ASR task. We follow the pipeline detailed in~\cite{Shi2022} to pre-process all
three datasets: For input audio features, we extract 26-dimensional log-filterbank outputs from raw audio sampled at 16 kHz with a 25 ms window and a frame shift of 10 ms, yielding $100$ audio frames per second. For video frames sampled at 25 Hz, we convert them to grayscale and crop them to a $96 \times 96$ region of interest (ROI) based on the face alignment, upon which a $88 \times 88$ ROI is cropped (randomly during training, centered during inference) as input to the neural network. \\

\textbf{Pre-trained models and fine-tuning for AV-ASR}: We use the English \texttt{Whisper} ASR~\cite{Radford2023} and \texttt{AV-HuBERT large}~\cite{Shi2022a} pre-trained on LRS3 and English subsets of Voxceleb2 datasets. For all experiments, the learning rate (LR) increases linearly to the peak LR within warmup steps, following a linear LR decrease to 0. The AV-ASR systems follow a two-stage fine-tuning pipeline: In the first stage, the \texttt{Whisper} ASR is fine-tuned on the LRS3 audio data for 90k steps (1000 warm-up steps, peak LR of $1.25\!\cdot\!10^{-5}$). In the second stage, the AV-ASR system is fine-tuned on the audiovisual data with a peak LR of $1.0\!\cdot\!10^{-4}$. The AV-ASR systems based on \texttt{Whisper tiny} and \texttt{Whisper base} are fine-tuned for 20k steps with 2000 warm-up steps. The AV-ASR systems based on \texttt{Whisper small} and \texttt{Whisper medium} are fine-tuned for 120k steps with 5000 warm-up steps. We augment the audio data with MUSAN babble noise with an SNR of 0dB~\cite{Snyder2015}. The AV-ASR systems based on \texttt{Whisper medium} are fine-tuned on a single \texttt{Nvidia H100}, other models are fine-tuned on a single \texttt{Nvidia A100}. \\
\textbf{Evaluation in noisy conditions}: First, we evaluate models' noise robustness in babble noise mixed by samples from the MUSAN dataset~\cite{Snyder2015} following previous work~\cite{Shi2022a}: \textit{Babble noise} is generated by mixing utterances from 30 different speakers from the MUSAN dataset~\cite{Snyder2015}, with each speaker exclusively assigned to either the training, validation, or test partition. Second, recent models~\cite{Yeo2025,Cappellazzo2025,Burchi2023,Ma2023} are trained and evaluated with babble noise from the NoiseX dataset~\cite{Varga1993}. For fair comparison under background noise, we also evaluate our models in the NoiseX babble noise, which is only seen during test in our experiments.

\subsection{Results and Discussion}

\input{table/ablation_results_avhubert_only}

In Table~\ref{tab:fusion_methods}, we compare \texttt{Whisper} ASR baselines (mode: A = audio) and AV-ASR systems (mode: AV = audiovisual) employing our proposed early fusion, the {\tt Flamingo}-based middle fusion, and our finally proposed dual-use fusion method. To validate the effectiveness of various fusion methods when scaling up the model size, we apply fusion methods to various sizes of \texttt{Whisper} ASR with a parameter count from 37M to 762M. Results are reported in clean condition and in MUSAN babble noise at a signal-to-noise ratio (SNR) of 0dB. \texttt{AV-HuBERT large} is used to extract visual features, already covering 325M parameters of each AV-ASR system reported in Table~\ref{tab:fusion_methods}.

In table segment \ding{192} based on \texttt{Whisper tiny} and table segment \ding{193} based on \texttt{Whisper base}, our early fusion presents the best performance in 0dB babble noise on the \texttt{dev} set, outperforming middle fusion~\cite{Rouditchenko2025} by 46\% relative (9.53\% vs.~17.54\%) in \ding{192} and by 51\% relative (8.38\% vs.~16.98\%) in \ding{193}. {\it Our dual-use method achieves the best or second best performance in all of the four {\tt dev/test} evaluation splits (clean and noisy).} Middle fusion~\cite{Rouditchenko2025} shows only sub-optimal noise robustness compared to the other two fusion methods. 

In table segment \ding{194} based on \texttt{Whisper small} and table segment \ding{195} based on \texttt{Whisper medium}, the early fusion method has a degradated performance especially in clean conditions, whereas our dual-use method achieves the best performance in both clean condition and 0dB babble noise on the \texttt{dev} set. In particular, compared to the middle fusion reference~\cite{Rouditchenko2025} in 0dB babble noise on the \texttt{test} set, \textit{our dual-use method demonstrates a \textbf{35\%} (\textbf{4.41\%} vs.~ 6.83\%) relative improvement in \ding{194} and a \textbf{57\%} (\textbf{4.07\%} vs.~9.53\%) relative improvement in \ding{195} at about the same parameter count. Our dual-use method consistently improves noise robustness across all four different model sizes \ding{192}-\ding{195}.} 

\input{table/ablation_modules_options.tex}

In Table~\ref{tab:ablation_modules}, we investigate two aspects of our proposed dual-use fusion approach: In the \texttt{Whisper} encoder, we compare the fusion design of addition vs.\ concatenation (feature concatenation following a linear fully connected layer $\mathrm{FC}(d)$). In the \texttt{AV-HuBERT large} encoder with 24 encoder blocks, we access the visual latent representation $\VEC{h}^\mathrm{V}_{1:\frac{T}{2}}$ after the given $\MAT{E}^\mathrm{V}()$ block \# to compute the visual feature sequence ${\VEC{v}}^\mathrm{V}_{1:T}$ by upsampling, projection with fully connected layer $\mathrm{FC}(d)$, and multiplication with a trainable scalar $\alpha$. The 0th $\MAT{E}^\mathrm{V}()$ block means visual feature sequence after the visual frontend $\MAT{G}^\mathrm{V}()$ and the positional encoding.
We decide for the best visual fusion option on the noisy \texttt{dev} set ("dual-use (ours)").

In the first row in Table~\ref{tab:ablation_modules}, the performance of the fine-tuned ASR baseline is presented. In the center table segment, the fusion design of addition excels that of concatenation significantly in all four evaluation splits, showing a 57\% (5.28\% vs.~12.15\%) relative improvement on the \texttt{dev} set in 0dB babble noise. {\it This reveals the importance of zero-initialized visual features ${\VEC{v}}^\mathrm{V}_{1:T}$, obtained by the trainable scalar $\alpha$ being zero-initialized in our additive visual fusion.} It enables a smooth fine-tuning start followed by gradually injecting visual features into the \texttt{Whisper} ASR. 

In the lower table segment, in 0dB babble noise on the \texttt{dev} set, we observe a monotonical performance improvement with visual latent representations from later encoder blocks (0th: 10.48\% $\!\to\!$ 24th: 5.28\%). The setting with additive fusion design and visual latent representations from the 24th \texttt{AV-HuBERT large} encoder block delivers the best WER (5.28\%) in 0dB babble noise on the \texttt{dev} set. Therefore, this setting is used for further experiments.

\input{table/benchmark.tex}

In Table~\ref{tab:benchmark}, we fine-tune our proposed dual-use method based on \texttt{Whisper small} and \texttt{Whisper medium} by scaling up the amount of fine-tuning data. Models are fine-tuned on 433h (LRS3), 654h (LRS3 and LRS2), 1929h (LRS3, LRS2, and Voxceleb-EN) audiovisual data. In addition, we compare our dual-use method with other models on the LRS3 AV-ASR benchmark. The WERs are reported on the LRS3 \texttt{test} set in clean conditions and two types of babble noise conditions at an SNR of -5dB, 0dB, and 5dB. The first type is babble noise from the MUSAN dataset~\cite{Snyder2015} in the upper table segment. The second type is babble noise from the NoiseX dataset~\cite{Varga1993} in the lower table segment. We also report the average WER over the clean and noisy conditions at these three SNRs. 

In the upper table segment evaluated in MUSAN babble noise, we first present three models on the LRS3 AV-ASR benchmark: \texttt{mWhisper Flamingo} applies {\tt Flamingo}-based middle fusion to \texttt{Whisper medium} ASR, which is fine-tuned on the MuAViC multilingual audiovisual dataset~\cite{Anwar2023} with 1141 hours of data. \texttt{AV-HuBERT}~\cite{Shi2022a} is built by a pre-trained \texttt{AV-HuBERT large} encoder and a random initialized decoder, which is fine-tuned on 433h of LRS3. \texttt{CMA}~\cite{Kim2024} contains a cross-modal attention after the audiovisual frontend to enrich video features, which is based on \texttt{AV-HuBERT} and fine-tuned on 1929h of data.

For our proposed dual-use method, by comparison among the same amount of fine-tuning data, the \texttt{Whisper medium} backbone provides a lower average WER compared to the \texttt{Whisper small} backbone, e.g., an 8.5\% relative reduction (4.08\% vs.~4.46\%) with 1929h of fine-tuning data. By comparison within the same \texttt{Whisper} backbone, increasing the amount of fine-tuning data leads to significant performance improvements, especially in clean conditions. For \texttt{Whisper medium}, e.g., by increasing the fine-tuning data from 433h to 1929h, we achieve a 19\% relative reduction on the average WER (5.06\% vs.~4.08\%), and a 28\% relative reduction in clean condition (1.59\% vs.~1.15\%). 

In the lower table segment evaluated in NoiseX babble noise, three models on the benchmark are exhibited. The first model is \texttt{Efficient Conformer}~\cite{Burchi2023}, which comprises 62M parameters and is trained from scratch on 818h of data. The \texttt{MMS-Llama}~\cite{Yeo2025} is a multi-modal LLM based on the \texttt{Llama3.2} 3B decoder, which is fine-tuned on LRS3 and Voxeleb-EN with 1759h of data. We downloaded their checkpoint and evaluated the performance. \texttt{Llama-AVASR} is also a multi-modal LLM based on the \texttt{Llama3.1} 8B decoder, which is fine-tuned on LRS3 and Voxceleb-EN as well. Leveraging the strong decoding ability of the LLM decoder by pre-training on 15 trillion text tokens~\cite{Grattafiori2024}, \texttt{MMS-Llama} and \texttt{Llama-AVASR} achieve the overall best WER (0.88\%) and the second best WER (0.90\%) in clean condition, respectively.

In the lower table segment, for our dual-use method, we evaluate the same models as shown in the upper table segment on the unseen NoiseX babble noise. We start from using our dual-use method on \texttt{Whisper small} and fine-tuning on 433h data, giving an average WER of 5.29\%. By applying our dual-use method to a larger model from \texttt{Whisper small} to \texttt{Whisper medium} (parameter count: 652M $\! \to\!$ 1390M), we obtain a 4\% (5.06\% vs.~5.29\%) relative reduction of the average WER. Next, by increasing the data size from 433h to 1929h, we achieve a 12\% (4.43\% vs.~5.06\%) relative reduction of the average WER.

{\it With the average WER of 4.08\% (MUSAN babble noise) and 4.43\% (NoiseX babble noise), we claim a new state of the art on the LRS3 AV-ASR benchmark with our dual-use method.}

\section{Conclusions}
In this work, we proposed a dual use of visual features (dual-use) both in \texttt{Whisper} encoder and decoder to build an AV-ASR system. This models audiovisual interactions by \texttt{Whisper} transformer encoder blocks, and delivers both modalities to the \texttt{Whisper} decoder for modality-aware decoding. Our proposed dual-use method consistently improves noise robustness of \texttt{Whisper}-based AV-ASR of various sizes. In addition, our dual-use method significantly outperforms the {\tt Flamingo}-based middle fusion method in noisy conditions at about the same parameter count, e.g., a 35\% relative improvement (4.41\% WER vs.~6.83\%) based on \texttt{Whisper small}, and a 57\% relative improvement (4.07\% WER vs.~9.53\%) in 0dB MUSAN babble noise. With audiovisual fine-tuning data of 1929 hours, our dual-use method achieves 4.08\% (MUSAN babble noise) and 4.43\% (NoiseX babble noise) average WER across various SNRs, \textit{claiming a new state-of-the-art in noisy conditions on the LRS3 AV-ASR benchmark.}
\label{sec:conclusions}

\vfill\pagebreak

\section{Acknowledgments}
\label{sec:ack}
\noindent The research leading to these results has received funding from the Bundesministerium für Forschung, Technologie und Raumfahrt (BMFTR) under funding code 03VP10991 (BesserLesen project).

\section{Compliance with Ethical Standards}
This research study was conducted retrospectively using human subject data made available in open access by~\cite{Afouras2018b,Chung2016,Chung2018}. Ethical approval was not required.

\bibliographystyle{IEEEbib}
\bibliography{ifn_spaml_bibliography}

\end{document}

%% file: macros/tikz_init.tex
\usepackage{tikz}
\usepackage{pgfplots}

\pgfplotsset{compat=newest}

\usetikzlibrary{arrows,automata}
\usetikzlibrary{shapes.geometric,shapes.arrows,decorations.pathmorphing}
\usetikzlibrary{matrix,chains,scopes,positioning,arrows,fit}
\usetikzlibrary{patterns}
\usetikzlibrary{chains,backgrounds,calc}

%% file: macros/mathmacros.tex
\newcommand{\VEC}[1]{\mathbf{#1}}          
\newcommand{\MAT}[1]{\mathbf{#1}}          

\newcommand{\putindex}[3]{\vtop{\hbox{\hspace{#3} $#1$}
            \hbox{\raise 6mm \hbox{$\scriptscriptstyle #2$}}}}

\newcommand{\gradx}[0]{\vtop{\hbox{\rm grad}
            \hbox{\raise 2.5mm \hbox{\rm \hspace{2mm} \footnotesize x}}}}

\newcommand{\grady}[0]{\vtop{\hbox{\rm grad}
            \hbox{\raise 2.5mm \hbox{\rm \hspace{2mm} \footnotesize y}}}}

\newcommand{\grad}[1]{\vtop{\hbox{\rm grad}
            \hbox{\raise 2.5mm \hbox{#1}}}}

\newcommand{\btb}{     \begin{tabbing}             }
\newcommand{\bte}{     \end{tabbing}               }

%% file: macros/tu_bs_colors.tex
\definecolor{tu0}{rgb}{0.7451, 0.1176, 0.2353}

\definecolor{tu1}{rgb}{1.0000, 0.8039, 0.0000}
\definecolor{tu11}{rgb}{1.0000, 0.8627, 0.3020}
\definecolor{tu12}{rgb}{1.0000, 0.9020, 0.4980}
\definecolor{tu13}{rgb}{1.0000, 0.9412, 0.6980}
\definecolor{tu14}{rgb}{1.0000, 0.9608, 0.8000}

\definecolor{tu2}{rgb}{0.9804, 0.4314, 0.0000}
\definecolor{tu21}{rgb}{0.9882, 0.6039, 0.3020}
\definecolor{tu22}{rgb}{0.9882, 0.7137, 0.4980}
\definecolor{tu23}{rgb}{0.9922, 0.8275, 0.6980}
\definecolor{tu24}{rgb}{0.9961, 0.8863, 0.8000}

\definecolor{tu3}{rgb}{0.6902, 0.0000, 0.2745}
\definecolor{tu31}{rgb}{0.7529, 0.2000, 0.4196}
\definecolor{tu32}{rgb}{0.8431, 0.4980, 0.6353}
\definecolor{tu33}{rgb}{0.9216, 0.7490, 0.8196}
\definecolor{tu34}{rgb}{0.9529, 0.8510, 0.8902}

\definecolor{tu4}{rgb}{0.4863, 0.8039, 0.9020}
\definecolor{tu41}{rgb}{0.6431, 0.8627, 0.9333}
\definecolor{tu42}{rgb}{0.7412, 0.9020, 0.9490}
\definecolor{tu43}{rgb}{0.8431, 0.9412, 0.9686}
\definecolor{tu44}{rgb}{0.8980, 0.9608, 0.9804}

\definecolor{tu5}{rgb}{0.0000, 0.5020, 0.7059}
\definecolor{tu51}{rgb}{0.3020, 0.6510, 0.7961}
\definecolor{tu52}{rgb}{0.5490, 0.7765, 0.8667}
\definecolor{tu53}{rgb}{0.7490, 0.8745, 0.9255}
\definecolor{tu54}{rgb}{0.8510, 0.9255, 0.9569}

\definecolor{tu6}{rgb}{0.0000, 0.3255, 0.4549}
\definecolor{tu61}{rgb}{0.2510, 0.4941, 0.5922}
\definecolor{tu62}{rgb}{0.5490, 0.6941, 0.7529}
\definecolor{tu63}{rgb}{0.7490, 0.8314, 0.8627}
\definecolor{tu64}{rgb}{0.8510, 0.8980, 0.9176}

\definecolor{tu7}{rgb}{0.7765, 0.9333, 0.0000}
\definecolor{tu71}{rgb}{0.8431, 0.9529, 0.3020}
\definecolor{tu72}{rgb}{0.8863, 0.9647, 0.4980}
\definecolor{tu73}{rgb}{0.9333, 0.9804, 0.6980}
\definecolor{tu74}{rgb}{0.9569, 0.9882, 0.8000}

\definecolor{tu8}{rgb}{0.5373, 0.6431, 0.0000}
\definecolor{tu81}{rgb}{0.6784, 0.7490, 0.3020}
\definecolor{tu82}{rgb}{0.7686, 0.8196, 0.4980}
\definecolor{tu83}{rgb}{0.8588, 0.8941, 0.6980}
\definecolor{tu84}{rgb}{0.9059, 0.9294, 0.8000}

\definecolor{tu9}{rgb}{0.0000, 0.4431, 0.3373}
\definecolor{tu91}{rgb}{0.3020, 0.6118, 0.5373}
\definecolor{tu92}{rgb}{0.5490, 0.7490, 0.7020}
\definecolor{tu93}{rgb}{0.7490, 0.8588, 0.8353}
\definecolor{tu94}{rgb}{0.8549, 0.9176, 0.9059}

\definecolor{tu10}{rgb}{0.8000, 0.0000, 0.6000}
\definecolor{tu101}{rgb}{0.8706, 0.3490, 0.7412}
\definecolor{tu102}{rgb}{0.9216, 0.6000, 0.8392}
\definecolor{tu103}{rgb}{0.9608, 0.8000, 0.9216}
\definecolor{tu104}{rgb}{0.9804, 0.8980, 0.9608}

\definecolor{tu110}{rgb}{0.4627, 0.0000, 0.4627}
\definecolor{tu111}{rgb}{0.5961, 0.2510, 0.5961}
\definecolor{tu112}{rgb}{0.7294, 0.4980, 0.7294}
\definecolor{tu113}{rgb}{0.8392, 0.6980, 0.8392}
\definecolor{tu114}{rgb}{0.9216, 0.8510, 0.9216}

\definecolor{tu120}{rgb}{0.4627, 0.0000, 0.3294}
\definecolor{tu121}{rgb}{0.6118, 0.3020, 0.5333}
\definecolor{tu122}{rgb}{0.7569, 0.5490, 0.6980}
\definecolor{tu123}{rgb}{0.8667, 0.7490, 0.8314}
\definecolor{tu124}{rgb}{0.9216, 0.8510, 0.9020}

\definecolor{tu130}{rgb}{0.0314, 0.0314, 0.0314}
\definecolor{tu131}{rgb}{0.3725, 0.3725, 0.3725}
\definecolor{tu132}{rgb}{0.5882, 0.5882, 0.5882}
\definecolor{tu133}{rgb}{0.7529, 0.7529, 0.7529}
\definecolor{tu134}{rgb}{0.8667, 0.8667, 0.8667}

%% file: figure/whisper_based_avasr_with_avhubert_encoder.tikz.tex
	\tikzset{%
		tipA/.tip={Triangle[angle=40:4pt]}
	}
	\tikzset{
		>=stealth',
		punktchain/.style={
			rectangle,rounded corners, 
			draw=black, thick,
			text width=7em, 
			minimum height=0.4cm, 
			text centered,fill=white, 
			on chain=a},
		line/.style={draw, thick, ->},
		every join/.style={->, thick,shorten >=1pt},
	}
	\tikzset{
		>=stealth',
		punkt_mult/.style={
			circle, 
			minimum size=0.1cm,inner sep=0.0mm, 
			draw=black, 
			text centered,fill=white, thick, 
			on chain},
		arrow/.style={draw, thick,-latex},
		every join/.style={->, thick,shorten >=1pt},
	}
	\tikzset{
		>=stealth',
		loet/.style={
			draw,circle,fill=black,inner sep=0.3mm,outer sep=0.0mm},
		arrow/.style={draw, thick,-tipA},
		every join/.style={->, thick,shorten >=1pt},
	}

	\begin{tikzpicture}[node distance=0.4cm,scale=1, every node/.style={scale=0.9}]
		
		{  [start chain=a going above]
			\def\x{-8mm}
			\def\sc{1.11}
			
			\node[on chain=a,outer sep=0mm, font=\fontsize{10pt}{10pt}\selectfont] (feats) {$\VEC{x}_{1:2T}^\mathrm{A}$};
			\node[punktchain,text width=7em,yshift=-0.0cm,xshift=\x*\sc, fill=red!0] (CONV) {Acous. Frontend \\ $\MAT{G}^\mathrm{A}()$};
			\node[punkt_mult,yshift=1mm,xshift=0] (enc_pos_emb) {$\bm{+}$};			
			\node[draw,circle, right=of enc_pos_emb, xshift=1mm,inner sep= 0mm, fill=white] (pos_enc) {\tikz \draw[x=0.95mm,y=1.3mm] (0,0) sin (1,1) cos (2,0) sin (3,-1) cos (4,0);};
			\node[draw=none, left=of enc_pos_emb, xshift=0,font=\fontsize{10pt}{10pt}\selectfont] (early_fusion_input){\colorbox{red!20}{${\VEC{v}}^\mathrm{V}_{1:T}$}};
			\node[right=of pos_enc,draw=none, yshift=0mm, text width=1.1cm,xshift=5mm,align=center] (pos_enc_label){};
			\draw[arrow] ( pos_enc) -- (enc_pos_emb);
			\draw[arrow] ( early_fusion_input) -- (enc_pos_emb);

			\node[punktchain,draw=black,fill=green!0](Encoder1)      {Encoder Block};
			\node[on chain=a,xshift=\x*\sc](dots){...};
			\node[right=of dots,xshift=-0.65cm, text width=1.8cm,align=center,yshift=1mm](dots_label){\scriptsize in total \vspace{-1mm} $N^\mathrm{ENC}$ \\ ENC blocks};
			
			\node[punktchain,draw=black,fill=green!0,xshift=-\x*\sc](Encoder2)      {Encoder Block};
			\node[left=of feats,draw=none, xshift=1.7cm,yshift=-0.20cm,text width=2.1cm](output_label_label)      {\footnotesize speech \\ \vspace{-0mm}  feature \\  \vspace{-1mm} sequence};
			\node[on chain=a,draw=none,xshift=1.55cm, yshift=0.0cm](out_node)      {};
			
			\draw[arrow] ( feats.north) --node[left,text=black!80,pos=0.3] {\scriptsize$B  \times 2T \times 80$} ([xshift=-\x] CONV.south);
			\draw[arrow] ( CONV.north) --node[left,text=black!80,pos=0.3] {\scriptsize$B\times T \times d$} (enc_pos_emb.south);
			\draw[arrow] ( enc_pos_emb.north) -- (Encoder1.south);
			\draw[arrow] ([xshift=\x] Encoder1.north) --node[right,text=black!80,pos=0.3] {\scriptsize$B\times T \times d$} ( dots.south);
			\draw[arrow] (dots.north) --node[left,text=black!50] {} ([xshift=\x] Encoder2.south);
			\draw[arrow,-] ([xshift=\x] Encoder2.north) |- node[right,text=black!80,pos=0.2] {\scriptsize$B\times T \times d$} node[left,text=black!80,pos=0.3,font=\fontsize{10pt}{10pt}\selectfont] {$\VEC{h}^\mathrm{AV}_{1:T}$} node[above,text=black!80,pos=0.92, yshift=3.4mm,font=\fontsize{10pt}{10pt}\selectfont] {$\VEC{h}^\mathrm{AV}_{1:T}$}  (out_node.south);
			
		}
		
		{  [start chain=b going above]
			\def\x{-8mm}
			\def\sc{1.11}
			
			\node[punktchain,on chain=b,right=of CONV,xshift=0.3cm,text width = 2.1cm] (embed) {Embedding};
			\node[draw=none,below=of embed,outer sep=0.5mm, yshift=-0.25cm,xshift=-\x*\sc,font=\fontsize{10pt}{10pt}\selectfont] (input_character) {${y}_{\ell\!-\!1}$};
			\node[punkt_mult,yshift=0mm,on chain=b,xshift=3.174cm, right=of enc_pos_emb] (enc_pos_emb) {$\bm{+}$};
			\node[draw,circle, left=of enc_pos_emb, xshift=2mm,inner sep= 0mm, fill=white] (pos_enc) {\tikz \draw[x=0.95mm,y=1.3mm] (0,0) sin (1,1) cos (2,0) sin (3,-1) cos (4,0);};
			\node[left=of pos_enc,draw=none, yshift=0mm, text width=1.1cm,xshift=5mm,align=center] (pos_enc_label){\footnotesize positional \\ \vspace{-1mm} encoding};
			\draw[arrow] ( pos_enc) -- (enc_pos_emb);
			
			\node[punktchain,on chain=b,draw=black,fill=orange!30,xshift=\x*\sc](Flamingo1)      {Flamingo Block};
			\node[punktchain,on chain=b,draw=black,fill=black!10, yshift=-0.15cm](Decoder1)      {Decoder Block};
			\node[on chain=b,xshift=\x*\sc](decoder_dots){...};
			\node[right=of decoder_dots,xshift=-0.65cm, text width=1.8cm,align=center,yshift=1mm](dots2_label){\scriptsize in total \vspace{-1mm} $2N^\mathrm{DEC}$ \\ DEC blocks};
			\node[punktchain,on chain=b,draw=black,fill=orange!30,xshift=-\x*\sc](Flamingo2)      {Flamingo Block};
			\node[punktchain,on chain=b,draw=black,fill=black!10,yshift=-0.15cm](Decoder2)      {Decoder Block};
			
			\node[punktchain,on chain=b,draw=black,text width=7.5em,rounded corners=0pt, yshift=0cm](output_fc)      {$\mathrm{FC}(D) \! + \! \text{Softmax}$};
			\node[on chain=b,draw=none,xshift=\x*\sc,yshift=0.8cm,font=\fontsize{10pt}{10pt}\selectfont](output_label)      {$\mathbf{P}_\ell$}; 
			\node[right=of output_label,draw=none, yshift=-0.0cm, xshift=-0.5cm,text width=2.1cm](output_label_label)      {\footnotesize output token \\ \vspace{-0.5mm} \footnotesize probabilities};
			\node[left=of input_character,draw=none, xshift=1.6cm,yshift=-0.12cm,text width=2.1cm](output_label_label)      {\footnotesize previous \\ input \\ \vspace{-1mm} token};
			
			\draw[arrow] (input_character.north) --node[left,text=black!80,pos=0.3] {\scriptsize$B \times 1 $} ([xshift=-\x]embed.south);
			\draw[arrow] ([xshift=-\x] embed.north) --node[left,text=black!80,pos=0.3] {\scriptsize$B \times 1 \times d$} (enc_pos_emb.south);
			\draw[arrow] ( enc_pos_emb.north) -- ([xshift=-\x] Flamingo1.south);
			\draw[arrow] ([xshift=\x] Flamingo1.north) --node[right,text=black!80,pos=0.4] {\scriptsize$B \times 1 \times d$} ([xshift=\x]Decoder1.south);
			\draw[arrow] ([xshift=\x] Decoder1.north) --node[right,text=black!80,pos=0.3] {\scriptsize$B \times 1 \times d$} (decoder_dots.south);
			\draw[arrow] ( decoder_dots.north) --node[right,text=black!50] {} ([xshift=\x] Flamingo2.south);
			\draw[arrow] ([xshift=\x]  Flamingo2.north) --node[right,text=black!80, pos=0.4] {\scriptsize$B \times 1 \times d$} ([xshift=\x] Decoder2.south);
			\draw[arrow] ([xshift=\x]  Decoder2.north) --node[right,text=black!80, pos=0.3] {\scriptsize$B \times 1 \times d$} ([xshift=\x] output_fc.south);
			
			\draw[arrow] ([xshift=\x] output_fc.north) --node[right,text=black!80,pos=0.1] {\scriptsize$B \times 1 \times D$} (output_label.south);
			
			\node[loet,left of=Decoder2,xshift=-1.185cm, yshift=-0.955cm] (dec_block_2_mag_in)      {};
			\node[loet,left of=decoder_dots,xshift=-0.298cm,yshift=1mm] (dec_block_mag_dot_in1) {} ;
			\draw[arrow] (dec_block_mag_dot_in1.east) -- ++(2.5mm,0em);
			\node[loet,left of=decoder_dots,xshift=-0.298cm,yshift=3mm] (dec_block_mag_dot_in2) {} ;
			\draw[arrow] (dec_block_mag_dot_in2.east) -- ++(2.5mm,0em);  
			\node[loet,left of=decoder_dots,xshift=-0.298cm,yshift=-1mm] (dec_block_mag_dot_in3) {} ;
			\draw[arrow] (dec_block_mag_dot_in3.east) -- ++(2.5mm,0em); 
			\node[loet,left of=decoder_dots,xshift=-0.298cm,yshift=-3mm] (dec_block_mag_dot_in3) {} ;
			\draw[arrow] (dec_block_mag_dot_in3.east) -- ++(2.5mm,0em); 
			
			\node[loet,right of=Flamingo2,xshift=1.13cm, yshift=-0.185cm] (dec_block_2_phase_in)      {};
			\node[loet,right of=decoder_dots,xshift=2.02cm,yshift=1mm] (dec_block_phase_dot_in1) {} ;
			\draw[arrow] (dec_block_phase_dot_in1.west) -- ++(-2.5mm,0em);
			\node[loet,right of=decoder_dots,xshift=2.02cm,yshift=3mm] (dec_block_phase_dot_in2) {} ;
			\draw[arrow] (dec_block_phase_dot_in2.west) -- ++(-2.5mm,0em); 
			\node[loet,right of=decoder_dots,xshift=2.02cm,yshift=-1mm] (dec_block_phase_dot_in3) {} ;
			\draw[arrow] (dec_block_phase_dot_in3.west) -- ++(-2.5mm,0em); 
			\node[loet,right of=decoder_dots,xshift=2.02cm,yshift=-3mm] (dec_block_phase_dot_in4) {} ;
			\draw[arrow] (dec_block_phase_dot_in4.west) -- ++(-2.5mm,0em); 
		}

		\draw[|-,arrow] ( out_node) |- node[right,text=black!50] {} (Decoder1.west);
		\draw[|-,arrow] ( out_node.south) |- node[right,text=black!50] {} (Decoder2.west);
		
		{  [start chain=c going above]
			\def\x{-8mm}
			\def\sc{1.11}
			\node[punktchain,on chain=c,right=of embed, xshift=0.3cm,text width=2.1cm, fill=red!0] (CONV2) {Visual Frontend \\ $\MAT{G}^\mathrm{V}()$};
			\node[yshift=0.0cm,below=of CONV2,xshift=-\x*\sc,font=\fontsize{10pt}{10pt}\selectfont] (feats2) {$\MAT{x}^\mathrm{V}_{1:\frac{T}{2}}$};
			
			\node[punkt_mult,yshift=1mm,on chain=c,xshift=-\x*\sc] (enc_pos_emb2) {$\bm{+}$};
			\node[draw,circle, left=of enc_pos_emb2, xshift=2mm,inner sep= 0mm, fill=white] (pos_enc) {\tikz \draw[x=0.95mm,y=1.3mm] (0,0) sin (1,1) cos (2,0) sin (3,-1) cos (4,0);};
			\node[left=of pos_enc,draw=none, yshift=0mm, text width=1.1cm,xshift=5mm,align=center] (pos_enc_label){\footnotesize positional \\ \vspace{-1mm} encoding};
			\draw[arrow] ( pos_enc) -- (enc_pos_emb2);

			\node[punktchain,on chain=c,draw=black,fill=green!0,xshift=\x*\sc](Encoder12)      {Encoder Block};
			\node[on chain=c,xshift=\x*\sc](dots2){...};
			\node[right=of dots2,xshift=-0.75cm, text width=1.8cm,align=center,yshift=1mm](dots2_label){\scriptsize in total 24 \\ \vspace{-1mm} ENC blocks};
			\node[left=of feats2,draw=none, xshift=1.6cm,yshift=0.00cm,text width=2.1cm](output_label_label)      {\footnotesize image \\ \vspace{-1mm} sequence};
			\node[punktchain,on chain=c,draw=black,fill=green!0,xshift=-\x*\sc](Encoder22)      {Encoder Block};
			
			\node[on chain=c,draw=none,xshift=-1.55cm, yshift=-0.01cm](out_node2)      {};
			
			\node[punktchain,on chain=c, draw=black,xshift=0, text width=8.0em, above=of Encoder22, yshift=1cm, fill=red!0, rounded corners=0pt](early_fusion)      {Upsample + $\mathrm{FC}(d)$};
			
			\node[on chain=c,punkt_mult,yshift=-0.2cm,xshift=\x*\sc] (scalar_early) {$\bm{\times}$};
			\node[draw=none, right=of scalar_early, xshift=-4.5mm,inner sep= 1mm, fill=white] (scalar_early_param) {$\alpha$};
			\node[draw=none, right=of scalar_early_param, xshift=-4.7mm,inner sep= 0mm, fill=white, text width=8em, yshift=0.19mm] (scalar_early_description) {\footnotesize trainable scalar};
			
			\node[on chain=c,draw=none,xshift=0,yshift=-0.2cm,font=\fontsize{10pt}{10pt}\selectfont](early_fusion_label)      {\colorbox{red!20}{${\VEC{v}}^\mathrm{V}_{1:T}$}};
			\node[draw=none,xshift=-0.55cm,yshift=-0.0cm, right= of early_fusion_label, text width=10em](early_fusion_description)      {\footnotesize input to \\  \vspace{-0.0cm} Whisper for\\ \vspace{-0.1cm}\footnotesize early fusion};
			
			\draw[arrow] ( feats2.north) --node[left,text=black!80,pos=0.3] {\scriptsize$B \! \times \! \frac{T}{2} \! \times \! 1 \! \times \! 88 \! \times \! 88$} ([xshift=-\x] CONV2.south);
			\draw[arrow] ([xshift=-\x] CONV2.north) --node[left,text=black!80,pos=0.3] {\scriptsize$B\times \frac{T}{2} \times 1024$} ( enc_pos_emb2.south);
			\draw[arrow] ( enc_pos_emb2.north) -- ([xshift=-\x] Encoder12.south);
			\draw[arrow] ([xshift=\x] Encoder12.north) --node[right,text=black!80,pos=0.4] {\scriptsize$B\times \frac{T}{2} \times 1024$} (dots2.south);
			\draw[arrow] ( dots2.north) --node[right,text=black!50] {} ([xshift=\x]  Encoder22.south);
			\draw[arrow,-] ( [xshift=\x] Encoder22.north) |- node[right,text=black!80,pos=0.2] {\scriptsize$B\times \frac{T}{2} \times 1024$} (out_node2.south);
			\draw[arrow] ( [xshift=\x] Encoder22.north) -- node[left,text=black!80,pos=0.5, xshift=0.1cm,yshift=0.1cm, font=\fontsize{10pt}{10pt}\selectfont] {$\VEC{h}^\mathrm{V}_{1:\frac{T}{2}}$} ([xshift=\x]early_fusion.south);
			\draw[arrow] ([xshift=\x]early_fusion.north) -- node[right,text=black!80,pos=0.5, xshift=0.0cm] {\scriptsize$B\times T \times d$} (scalar_early.south);
			\draw[arrow] (scalar_early.north) -- node[right,text=black!80,pos=0.5] {\scriptsize$B\times T \times d$} (early_fusion_label.south);
			
			\draw[|-,arrow] ( out_node2) |- node[right,text=black!50] {} (Flamingo1.east);
			\draw[|-,arrow] ( out_node2.south) |- node[right,text=black!50] {} (Flamingo2.east);

		}
		
		\node[above=of Encoder2,anchor=south, xshift=0cm,yshift=-0.07cm,align=right] () {\footnotesize \bf Trans. Enc. $\MAT{E}^\mathrm{A}()$};
		\node[above=of Encoder2,anchor=south, xshift=-0.3cm,yshift=2.3cm,align=left] () {\footnotesize \texttt{\textbf{Whisper}} \textbf{ASR} \\ \footnotesize \bf based on Acoustics};
		\node[above=of Encoder22,anchor=south, xshift=0.45cm,yshift=0.45cm,align=left] () {\footnotesize \bf AV-HuBERT};
		\node[above=of Encoder22,anchor=south, xshift=0.3cm,yshift=-0.07cm,align=left] () {\footnotesize \bf Trans. Enc. $\MAT{E}^\mathrm{V}()$};
		\node[above=of output_fc,anchor=south, xshift=0.4cm,yshift=-0.15cm,align=right] () {\footnotesize\bf Decoder $\MAT{D}()$ };
		

		\begin{pgfonlayer}{background}
			\path (output_fc.east |- output_fc.north)+(0.15,0.85) node (g) {};
			\path (CONV.west |- CONV.south)+(-0.450,-0.18) node (h) {};
			\path[fill=blue!10,rounded corners, draw=black!50,draw=none]
			(g) rectangle (h);
	
		\begin{pgfonlayer}{background}
			\coordinate (bgNW) at ($(early_fusion_input.north west)+(-1.3pt,  0.05pt)$);
			\coordinate (bgSE) at ($(early_fusion_input.south east)+( 3.2pt,  -0.3pt)$);
			\node[fill=white, rounded corners=0pt, inner sep=0pt,
			fit=(bgNW) (bgSE)] (early_fusion_bg) {};
		\end{pgfonlayer}
		
		\begin{pgfonlayer}{background}
			\coordinate (bgNW) at ($(early_fusion_input.north west)+(-0.0pt,  0.05pt)$);
			\coordinate (bgSE) at ($(early_fusion_input.south east)+( 13.2pt,  -0.3pt)$);
			\node[fill=white, rounded corners, inner sep=0pt,
			fit=(bgNW) (bgSE)] (early_fusion_bg) {};
		\end{pgfonlayer}
		
		\end{pgfonlayer}
		
		\begin{pgfonlayer}{background}
			\path (Encoder22.west |- Encoder22.north)+(-0.2,1.20) node (g) {};
			\path (CONV2.east |- CONV2.south)+(0.3,-0.2) node (h) {};
			\path[fill=yellow!15,rounded corners, draw=black!50,draw=none]
			(g) rectangle (h);
		\end{pgfonlayer}

		\begin{pgfonlayer}{background}
			\path (Encoder2.west |- Encoder2.north)+(-0.15,0.8) node (g) {};
			\path (Encoder1.east |- Encoder1.south)+(0.15,-0.2) node (h) {};
			\path[fill=green!30,rounded corners, draw=black!50,draw=none]
			(g) rectangle (h);
		\end{pgfonlayer}
		
		\begin{pgfonlayer}{background}
			\path (Encoder22.west |- Encoder22.north)+(-0.1,0.8) node (g) {};
			\path (Encoder12.east |- Encoder12.south)+(0.15,-0.2) node (h) {};
			\path[fill=yellow!40,rounded corners, draw=black!50,draw=none]
			(g) rectangle (h);
		\end{pgfonlayer}
		
		\begin{pgfonlayer}{background}
			\path (output_fc.west |- output_fc.north)+(-0.15,0.7) node (g) {};
			\path (embed.east |- embed.south)+(0.150,-0.18) node (h) {};
			\path[fill=blue!20,rounded corners, draw=black!50,draw=none]
			(g) rectangle (h);
			
		\end{pgfonlayer}

	\end{tikzpicture}

%% file: table/ablation_results_avhubert_only.tex
\usetikzlibrary{backgrounds} 
\newcommand{\whiteinsideDing}[2][0.35em]{
	\tikz[baseline=(d.base)]{
		\node[inner sep=0pt, outer sep=0pt] (d) {\ding{#2}};
		\begin{scope}[on background layer]
			\fill[white] (d.center) circle[radius=#1]; 
		\end{scope}
	}%
}
\begin{table}[t!] 
	\centering
	\caption{WER (\%) of {\tt Whisper} ASR (mode: A) and {\tt Whisper}-based AV-ASR (mode: AV) with \textbf{various fusion methods}. Fusion methods are applied to \textbf{various sizes of \texttt{Whisper} ASR}. Results are reported on the LRS3 {\tt dev} and {\tt test} sets in clean condition and in babble noise with an SNR of 0dB. Best results in each table segment are in bold, second best are \underline{underlined}. }\label{tab:fusion_methods}
	\resizebox{\linewidth}{!}{
		\fontsize{12}{14}\selectfont
		\begin{tabular}{@{\hskip 0.05cm} l @{\hskip 0.20cm} c  c	rr 	rr @{\hskip 0.05cm}}
			\toprule
			\multirow{3}[2]{*} { \shortstack[c]{Method}}  & \multirow{3}[2]{*}{\shortstack[c]{\vspace{0.09cm} Mode}} & %
			\multirow{3}[2]{*} {\shortstack[c]{\vspace{0.09cm} \# Par. \\ (M)}} & \multicolumn{4}{c@{\hskip 0.0em}}{WER (\%) on LRS3}   \\
			\cmidrule(lr{0.3em}){4-7}
			& & & \multicolumn{2}{c@{\hskip 0.0em}}{Clean} & \multicolumn{2}{c@{\hskip 0.0em}}{0dB Babble} \\
			\cmidrule(lr{0.3em}){4-5}	\cmidrule(lr{0.3em}){6-7}		
			& &	& \tt dev	&	\tt test & \tt dev	& \tt test \\
			\cmidrule(lr{0.3em}){1-7}
			
			\rowcolor{gray!20}
			\multicolumn{7}{@{\hskip 0.9em}c@{\hskip 0.8em}}{\whiteinsideDing{192} \textit{{\tt Whisper tiny}, 20k updates}} \\	
			
			{\hskip 0.05cm} ASR zero-shot	& A &	37	&	12.21	&	3.26 	\hskip 0.06cm	&	59.20	&	40.82 	\hskip 0.05cm \\	
			{\hskip 0.05cm} ASR fine-tuned	& A &	37	&	6.96	&	2.64	\hskip 0.06cm	&	33.38	&	19.58	\hskip 0.05cm \\
			
			\cmidrule(lr{0.55em}){2-7}
			
			{\hskip 0.05cm} Early fusion (ours) & AV			&	363	&	\textbf{5.77}	&	\underline{2.41} \hskip 0.06cm	&	\textbf{9.53}	&	\underline{8.45}	\hskip 0.05cm \\
			
			
			{\hskip 0.05cm} Middle fusion~\cite{Rouditchenko2025} & AV			&	370	&	6.86	&	2.64 \hskip 0.06cm	&	17.54	&	10.86	\hskip 0.05cm \\
			
			{\hskip 0.05cm} Dual-use (ours) & AV			&	370	&	\underline{6.20}	&	\textbf{2.04}	\hskip 0.06cm	&	\underline{10.86}	&	\textbf{8.17}	\hskip 0.05cm \\
			
			\rowcolor{gray!20}
			\multicolumn{7}{@{\hskip 0.9em}c@{\hskip 0.8em}}{\whiteinsideDing{193} \textit{{\tt Whisper base}, 20k updates}} \\	
			
			{\hskip 0.05cm} ASR zero-shot	& A &	72	&	7.89	&	2.50	\hskip 0.06cm	&	49.89	&	29.28	\hskip 0.05cm \\	
			{\hskip 0.05cm} ASR fine-tuned	& A &	72	&	5.07	&	1.77	\hskip 0.06cm	&	25.70	&	14.16	\hskip 0.05cm \\
			
			\cmidrule(lr{0.55em}){2-7}
			{\hskip 0.05cm} Early fusion (ours) & AV			&	398	&	6.25	&	1.79	\hskip 0.06cm	&	\textbf{8.38}	&	\underline{6.91}	\hskip 0.05cm \\
			
			{\hskip 0.05cm} Middle fusion~\cite{Rouditchenko2025} & AV			&	417	&	\underline{5.01}	&	\underline{1.67}	\hskip 0.06cm 	&	16.98	&	10.13	\hskip 0.05cm \\
			
			{\hskip 0.05cm} Dual-use (ours)& AV			&	417	&	\textbf{4.79}	&	\textbf{1.54}	\hskip 0.06cm	&	\underline{8.88}	&	\textbf{6.23}	\hskip 0.05cm \\
			
			\rowcolor{gray!20}
			\multicolumn{7}{@{\hskip 0.9em}c@{\hskip 0.8em}}{\whiteinsideDing{194} \textit{{\tt Whisper small}, 120k updates}} \\	
			
			{\hskip 0.05cm} ASR zero-shot	& A &	240	&	7.26	&	2.20	\hskip 0.06cm	&	42.49	&	22.06	\hskip 0.05cm \\	
			{\hskip 0.05cm} ASR fine-tuned	& A &	240	&	3.45	&	\textbf{1.27}	\hskip 0.06cm	&	19.83	&	10.49	\hskip 0.05cm \\
			
			\cmidrule(lr{0.55em}){2-7}
			{\hskip 0.05cm} Early fusion (ours) & AV			&	566	&	23.20	&	36.22	\hskip 0.06cm	&	23.17	&	36.99	\hskip 0.05cm	\\
			
			{\hskip 0.05cm} Middle fusion~\cite{Rouditchenko2025} & AV			&	652	&	\underline{3.29}	&	\underline{1.49}	\hskip 0.06cm	&	\underline{9.23}	&	\underline{6.83}	\hskip 0.05cm	\\
			
			{\hskip 0.05cm} Dual-use (ours) & AV			&	652	&	\textbf{3.24}	&	1.60	\hskip 0.06cm	&	\textbf{5.28}	&	\textbf{4.41}	\hskip 0.05cm	\\
			
			\rowcolor{gray!20}
			\multicolumn{7}{@{\hskip 0.9em}c@{\hskip 0.8em}}{\whiteinsideDing{195} \textit{{\tt Whisper medium}, 120k updates}} \\	
			
			{\hskip 0.05cm} ASR zero-shot	& A &	762	&	7.38	&	2.39	\hskip 0.06cm	&	32.10	&	16.99	\hskip 0.05cm	\\	
			{\hskip 0.05cm} ASR fine-tuned	& A &	762	&	3.24	&	\textbf{1.20}	\hskip 0.06cm	&	17.00	&	\underline{8.90}	\hskip 0.05cm	\\
			
			\cmidrule(lr{0.55em}){2-7}
			{\hskip 0.05cm} Early fusion (ours) & AV			&	1089	&	11.32	&	14.08	\hskip 0.06cm	&	\underline{13.33}	&	18.46	\hskip 0.05cm	\\
			
			{\hskip 0.05cm} Middle fusion~\cite{Rouditchenko2025}	& AV		&	1391	&	\underline{3.05}	&	\underline{1.58}	\hskip 0.06cm	&	16.70	&	9.53	\hskip 0.05cm	\\
			
			{\hskip 0.05cm} Dual-use (ours) & AV			&	1391	&	\textbf{2.80}	&	1.59	\hskip 0.06cm	&	\textbf{5.13}	&	\textbf{4.07}	\hskip 0.05cm	\\
			
			\bottomrule
	\end{tabular}}
\end{table}

%% file: table/ablation_modules_options.tex
\begin{table}[t!]
	\centering
	\caption{WER (\%) of \textbf{our proposed dual-use \texttt{Whisper small} method} with visual latent representation $\VEC{h}^\mathrm{V}_{1:\frac{T}{2}}$ options from various blocks of {\tt AV-HuBERT large} and encoder visual fusion design (addition or concatenation). Results are reported on the LRS3 {\tt dev} and {\tt test} sets in clean condition and in babble noise with an SNR of 0dB. Best results are in bold, second best are \underline{underlined}.}
	\label{tab:ablation_modules}
	\resizebox{\linewidth}{!}{
		\fontsize{12}{14}\selectfont
		\begin{tabular}{@{\hskip 0.0cm} l @{\hskip 0.20cm} c	c	c	rr	rr}
			\toprule
			\multirow{3}[2]{*} {\shortstack[c]{\hskip 0.12cm Method}}  & \multicolumn{2}{c@{\hskip 0.3em}}{Visual fusion} %
			& \multirow{3}[2]{*} {\shortstack[c]{\vspace{0.09cm} \# Par. \\ (M)}} & \multicolumn{4}{c@{\hskip 0.3em}}{WER (\%) on LRS3}   \\
			\cmidrule(lr{0.3em}){2-3} \cmidrule(lr{0.3em}){5-8}
			& \multirow{2}[2]{*}{\shortstack[c]{\vspace{0.09cm} Fusion \\ design }} & \multirow{2}[2]{*} {\shortstack[c]{\vspace{0.03cm} $\VEC{E}^\mathrm{V}()$ \\ \vspace{0.12cm} block}} &	& \multicolumn{2}{c@{\hskip 0.3em}}{Clean} & \multicolumn{2}{c@{\hskip 0.3em}}{0dB Babble} \\
			\cmidrule(lr{0.3em}){5-6}	\cmidrule(lr{0.3em}){7-8}		
			&  &	& 	& \l{\fontsize{12}{14}\selectfont dev}	&	\l{\fontsize{12}{14}\selectfont test} & \l{\fontsize{12}{14}\selectfont dev}	&	\l{\fontsize{12}{14}\selectfont test}		\\	
			\cmidrule(lr{0.3em}){1-8}
			
			\multicolumn{2}{@{\hskip -1.3cm}c}{ASR fine-tuned} && 240 & 3.45 & \textbf{1.27} & 19.83 & 10.49 \\

			\cmidrule(lr{0.3em}){1-8}
			{\hskip 0.05cm} Dual-use (ours) & add			&	24th	&	652	&	3.24	&	1.60	&	\textbf{5.28}	&	\underline{4.41}	\\
			{\hskip 0.05cm} ... & concat			&	24th	&	653	&	12.20	&	32.07	&	12.15	&	31.98	\\
			\cmidrule(lr{0.3em}){1-8}
			{\hskip 0.05cm} ... & add			&	16th	&	551	&	3.16	&	1.46	&	\underline{5.99}	&	\textbf{4.28}	\\
			{\hskip 0.05cm} ... & add			&	12th	&	501	&	\textbf{2.99}	&	\underline{1.45}	&	6.03	&	4.65	\\
			{\hskip 0.05cm} ... & add			&	8th		&	450	&	3.30	&	1.53	&	6.66	&	4.81	\\
			{\hskip 0.05cm} ... & add			&	4th	&	400	&	3.31	&	1.60	&	7.62	&	5.37	\\
			{\hskip 0.05cm} ... & add			&	0th	&	\textbf{337}	&	\underline{3.15}	&	1.71	&	10.48	&	6.49	\\
			
			\bottomrule
	\end{tabular}
}
	
\end{table}

%% file: table/benchmark.tex
\begin{table}[!t]
	\centering
	\caption{\textbf{LRS3 Benchmark on the AV-ASR task}. Models are evaluated on babble noise mixed by MUSAN samples or babble noise from the NoiseX dataset. Best results in each noisy condition are in bold, second best are \underline{underlined}. Only papers with number results in the following babble noise conditions or with open-source models are listed here. * Models are trained and evaluated with the same babble noise samples. \dag Numbers are cited from the respective paper.}
	\label{tab:benchmark}
	\resizebox{\columnwidth}{!}{%
		\begin{tabular}{@{}l  r@{\hspace{7pt}} r  r  r  r  r  r@{}}
			\toprule
			\multirow{3}{*}{Methods}
			& \multirow{3}{*}{\shortstack[c]{\#Par \\ (M)} }
			& \multirow{3}{*}{\shortstack[c]{Fine-\\tune \\ data (h)} }
			& \multicolumn{5}{c}{WER (\%) on LRS3 {\tt test} set} \\
			\cmidrule(lr){4-8}
			& & & \multicolumn{3}{c}{SNR} & \multirow{2}{*}{clean} & \multirow{2}{*}{avg.}\\
			 \cmidrule(lr){4-6}
			& & & -5dB & 0dB & 5dB &  &  \\
			\midrule
			\rowcolor{gray!20}
			\multicolumn{8}{c}{\textit{Babble noise mixed by MUSAN~\cite{Snyder2015} samples}} \\	
				
			\texttt{mWhisper Flam.}~\cite{Rouditchenko2025}\dag
			& 1390	& 1141 \hskipt
			& 27.20 & 8.70 & 4.80 & - & - \\
			
			\texttt{AV-HuBERT}~\cite{Shi2022a}\dag
			& 477	& 433 \hskipt
			& 16.60 & 5.80 & 2.60 & 1.40 & 6.60 \\
			
			\texttt{CMA}~\cite{Kim2024}\dag
			&	500	& 1929 \hskipt 
			& 11.90 & 4.40 & 2.40 & 1.50 & 5.05 \\
			
			\cmidrule{2-8}
			
			\multirow{3}{*}{\shortstack[c]{Dual-use (ours) \\ (\texttt{Whisper small})}}
			&	652	& 433 \hskipt & 12.36 & 4.41 & 2.23 & 1.60 & 5.15 \\
			&	652	& 654 \hskipt & 12.31 & 4.21 & 2.22 & 1.44 &  5.05 \\
			&	652	& 1929 \hskipt & \underline{11.34} & \underline{3.57} & \textbf{1.72} & \underline{1.19} & \underline{4.46} \\

			\cmidrule{2-8}
			
			\multirow{3}{*}{\shortstack[c]{Dual-use (ours) \\ (\texttt{Whisper medium})}}
			&	1390	& 433 \hskipt& 12.47 & 4.07 & 2.12 & 1.59 & 5.06 \\
			&	1390	& 654 \hskipt& 11.93 & 3.70 & 2.06 & 1.43 & 4.78 \\
			&	1390	& 1929 \hskipt& \textbf{10.15} & \textbf{3.17} & \underline{1.85} & \textbf{1.15} & \textbf{4.08} \\

			\midrule
			\rowcolor{gray!20}
			\multicolumn{8}{c}{\textit{Babble noise from the NoiseX~\cite{Varga1993} dataset}} \\
			\texttt{Eff.\ Conf.}~\cite{Burchi2023} *\dag
			&	62	& 818	\hskipt
			& \textbf{11.20} &  4.90 &  3.10 & 1.80 & 5.25	\\
			
			\texttt{MMS-Llama}~\cite{Yeo2025} *
			&	$>$3000	& 1759 \hskipt
			& 42.58 &  9.70 &  3.11 & \textbf{0.88} & 14.07	\\
			
			
			\texttt{Llama-AVASR}\cite{Cappellazzo2025} *\dag
			&	$>$8000	& 1756 \hskipt
			& 16.40 &  4.20 &  2.30 & \underline{0.90} & 5.95	\\
			
			\cmidrule{2-8}
			
			\multirow{3}{*}{\shortstack[c]{Dual-use (ours) \\ (\texttt{Whisper small})}}
			&	652	& 433 \hskipt & 12.92 & 4.31 & 2.33 & 1.60 & 5.29 \\
			&	652	& 654 \hskipt & 12.88 & 3.94 & 2.14 & 1.44 & 5.10 \\
			&	652	& 1929 \hskipt & 11.94 & 3.64 & \textbf{1.71} & 1.20 & \underline{4.62} \\
			
			\cmidrule{2-8}
			
			\multirow{3}{*}{\shortstack[c]{Dual-use (ours) \\ (\texttt{Whisper medium})}}
			&	1390	& 433 \hskipt & 12.62 & 3.90 & 2.13 & 1.59 & 5.06 \\
			&	1390	& 654 \hskipt & 11.94 & \underline{3.61} & 2.00 & 1.43 & 4.74 \\
			&	1390	& 1929 \hskipt & \underline{11.27} & \textbf{3.41}& \underline{1.90} & 1.15 & \textbf{4.43}\\

			\bottomrule
		\end{tabular}%
	}
\end{table}